\documentclass[a4paper,11pt]{article}
\usepackage{pos}

\title{New dark matter production mechanism and the gravitational wave signals}

\author*{Fa Peng Huang}

\affiliation{MOE Key Laboratory of TianQin Mission, TianQin Research Center for
	Gravitational Physics \& School of Physics and Astronomy, Frontiers
	Science Center for TianQin, Gravitational Wave Research Center of CNSA, 
	Sun Yat-sen University,\\
Zhuhai 519082, China}

\emailAdd{huangfp8@sysu.edu.cn}

\abstract{The microscopic origin and production mechanism of dark matter (DM) remain central questions in cosmology and particle physics. While thermal freeze-out has long dominated DM model building, alternative non-thermal scenarios are gaining prominence. In this work, we explore novel production channels for heavy DM candidates, including pseudo-Nambu-Goldstone bosons (pNGBs), Q-balls, and filtered DM arising from early-universe phenomena such as primordial black hole (PBH) evaporation, superradiance, and first-order phase transitions (FOPTs). We demonstrate that these mechanisms naturally generate gravitational wave (GW) signals detectable by future observatories, such as  LISA, TianQin, Taiji, and Cosmic Explorer. This multi-messenger approach offers a promising pathway to probe the origin and nature of DM beyond conventional paradigms.}

\FullConference{19th International Conference on Topics in Astroparticle and Underground Physics (TAUP2025)\\
24-30 Aug 2025\\
Xichang, China\\}

\begin{document}
\maketitle

\section{Introduction}

The nature of dark matter (DM)---its microscopic identity and cosmological origin---remains one of the most profound mysteries in modern physics. Traditional models often rely on thermal freeze-out to explain the observed relic abundance, but null results from direct detection and collider experiments have motivated the exploration of alternative production mechanisms. In particular, macroscopic and non-thermal DM candidates, such as Q-balls, pseudo-Nambu--Goldstone bosons (pNGBs), and filtered DM, offer compelling possibilities rooted in early-universe dynamics. This work investigates two broad classes of non-thermal DM production:
\begin{itemize}
	\item \textbf{Primordial black hole (PBH) evaporation and superradiance}, which can emit pNGB DM and generate distinctive gravitational wave (GW) signals~\cite{Jiang:2025blz,Huang:2025hos}.
	\item \textbf{First-order phase transitions (FOPTs)}, which can produce Q-ball and gauged Q-ball DM, as well as filtered DM, accompanied by stochastic GW backgrounds~\cite{Huang:2017kzu,Jiang:2023nkj, Jiang:2023qbm,Jiang:2024zrb,Qiu:2025tmn,Jiang:2025xln}	.
\end{itemize}

These mechanisms not only yield viable DM relic densities but also produce GW signatures accessible to a wide range of detectors, including LISA, TianQin, Taiji,  and Cosmic Explorer. By correlating DM production with GW observables, we aim to establish a multi-messenger framework that connects cosmology, particle physics, and GW astronomy.

\section{pNGB dark matter from Primordial black hole  Hawking radiation and superradiance with its gravitational wave  signals}

In many well-motivated new physics models, the pNGB from U(1) symmetry breaking emerges as a promising DM candidate.  Meanwhile, pNGB DM candidate naturally suppress the direct detection signals.
We consider the Minimal pNGB DM model with explicit symmetry breaking term
to give DM mass based on the following potential
\begin{equation}
	V(H, S)=-\frac{\mu_H^2}{2}|H|^2+\frac{\lambda_H}{2}|H|^4-\frac{\mu_S^2}{2}|S|^2+\frac{\lambda_S}{2}|S|^4+\lambda_{H S}|H|^2|S|^2-\frac{m^2}{4}\left(S^2+S^{* 2}\right)
\end{equation}
The pNGB coupling is small due to the suppression of high symmetry-breaking scale, then how to produce enough DM relic density?

\begin{figure}[h]
	\begin{minipage}{0.465\linewidth}
		\vspace{3pt}
		\centerline{\includegraphics[width=\textwidth]{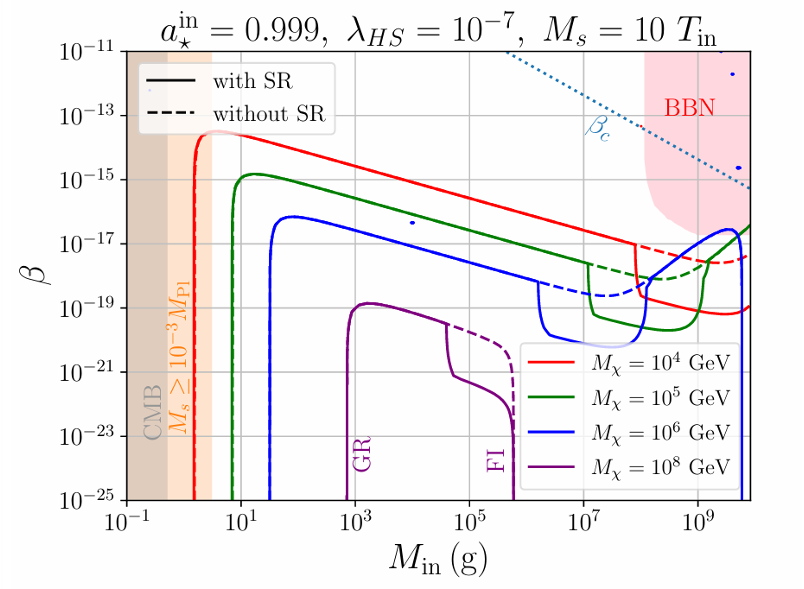}}
	\end{minipage}
	\begin{minipage}{0.465\linewidth}
		\vspace{3pt}
		\centerline{\includegraphics[width=\textwidth]{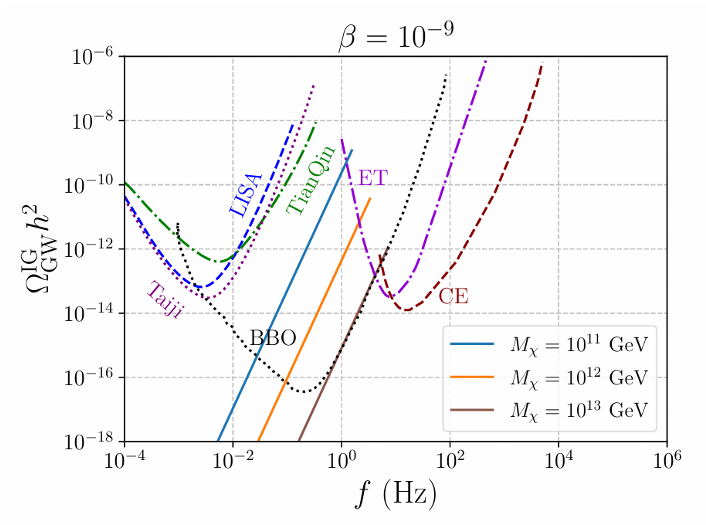}}
	\end{minipage}
	\caption{DM  from Hawking radiation and superradiance of PBH and the associated GWs. Left:  Contour plot of the initial fraction and mass of PBH that satisfy the correct DM abundance.  Right: Induced GWs from the DM production~\cite{Jiang:2025blz}.}
	\label{ppp}
\end{figure}

PBH can radiate  DM and  particles lighter than the Hawking temperature with the spectrum
\begin{equation}
	\frac{\mathrm{d}^2 N_i}{\mathrm{~d} p \mathrm{~d} t}=\frac{g_i}{2 \pi^2} \sum_{l=s_i} \sum_{m=-l}^l \frac{\sigma_{s_i}^{l m}\left(M_{\mathrm{PBH}}, p, a_{\star}\right)}{\exp \left[\left(E_i(p)-m \Omega\right) / T_{\mathrm{PBH}}\right]-(-1)^{2 s_i}} \frac{p^3}{E_i(p)}
\end{equation}

The Hawking radiation and superradiance of PBH can efficiently produce enough DM relic density as shown in the left panel of Fig.~\ref{ppp}.
The DM production process in the early universe can induce various GW signals and we show the induced GW spectra in the right panel of  Fig.~\ref{ppp}.

\section{Dark matter from first-order phase transition and gravitational wave}

Recently, dynamical DM formed by phase transition has became a new idea for heavy DM. Bubble wall in FOPT can be the “filter” to obtain the needed heavy DM when avoiding the unitarity constraints.  This DM production process is like the coffee-making process as shown in Tab.~\ref{coffee}.

\begin{table}[t]
	\centering

\begin{tabular}{|l|l|}
	\hline FOPT in the early universe & Coffee-making process \\
	\hline Bubble wall & filter \\
	\hline Case I:  (gauged) Q-ball DM & Large coffee beans \\
	\hline Case II:   filtered DM & Coffee \\
	\hline Phase transition GW & Aroma \\
	\hline
\end{tabular}
\caption{The DM production process and the coffee-making process }\label{coffee}
\end{table}

\subsection{Q-ball and gauged Q-ball dark matter}

Q-ball is the most typical non-topological soliton proposed by  Tsung-Dao Lee and  Sidney Coleman.  A spherically symmetric extended body that forms a non-topological soliton structure with a conserved global quantum number Q is called a Q-ball.  The expanding bubbles can help to produce Q-ball~\cite{Huang:2017kzu,Jiang:2023qbm,Jiang:2025xln}  or
gauged Q-ball~\cite{Jiang:2024zrb} DM with associated phase transition GW or braking GW~\cite{Qiu:2025tmn}. The DM relic density is sensitive to the FOPT dynamics. For example,
the gauged Q-ball DM density is determined by 
\begin{equation}
		\Omega_{\mathrm{Q}} h_{100}^2 
		\simeq 2.81 \times \left(\frac{s_0h_{100}^2}{\rho_c}\right) \left(\frac{\Gamma(T_\star)}{v_w}\right)^{3/16}s_\star^{-1/4}(F_\phi^{\mathrm{trap}}\eta_\phi)^{3/4} \lambda_h^{1/4}v_0\left(1+\frac{108^{1/4}\tilde{g}^2F_{\phi}^{\mathrm{trap}}\eta_{\phi}s_\star v_w^{3/4}}{5.4\pi^{7/4}\Gamma(T_\star)^{3/4}}\right) \nonumber
\end{equation}

\subsection{Hydrodynamic effects in Filtered dark matter}

We investigate the impact of hydrodynamic effects on the DM relic density. Through detailed calculations, we show that hydrodynamic modes, together with the associated heating processes, play a crucial role in shaping the final DM abundance. Moreover, the GWs generated during the corresponding phase transition provide a distinctive observational signature, offering a promising avenue to probe this new production mechanism~\cite{Jiang:2023nkj} with the following  DM relic density 
\begin{equation}
	\Omega_{\mathrm{DM}}^{(\mathrm{hy})} h^2=\frac{m_\chi^{\mathrm{in}}\left(n_\chi^{\mathrm{in}}+n_{\bar{\chi}}^{\mathrm{in}}\right)}{\rho_c / h^2} \frac{g_{\star 0} T_0^3}{g_{\star}\left(T_{-}\right) T_{-}^3} \simeq 6.29 \times 10^8 \frac{m_\chi^{\mathrm{in}}}{\mathrm{GeV}} \frac{\left(n_\chi^{\mathrm{in}}+n_{\bar{\chi}}^{\mathrm{in}}\right)}{g_{\star}\left(T_{-}\right) T_{-}^3}
\end{equation}
with
\begin{equation}
	n_\chi^{\text {in }} \simeq \frac{g_\chi T_{+}^3}{\gamma_w v_w}\left(\frac{\tilde{\gamma}_{+}\left(1-\tilde{v}_{+}\right) m_\chi^{\text {in }} / T_{+}+1}{4 \pi^2 \tilde{\gamma}_{+}^3\left(1-\tilde{v}_{+}\right)^2}\right) e^{-\frac{\tilde{\gamma}_{+}\left(1-\tilde{v}_{+}\right) m_\chi^{\text {in }}}{T_{+}}}
\end{equation}

\section{Conclusions and discussions}

We have presented various frameworks for exploring new DM production mechanisms beyond thermal freeze-out, focusing on early-universe processes such as PBH evaporation, superradiance, and FOPTs. These scenarios naturally give rise to heavy DM candidates including pNGBs, Q-balls, and filtered DM whose relic densities are set by non-thermal dynamics.
Crucially, these mechanisms also produce GW signals that serve as indirect probes of DM physics. We have outlined the expected GW signatures and their detectability across a broad spectrum of current and future observatories. This multi-messenger strategy not only expands the landscape of viable DM models but also provides observational handles to test them in the coming decade. Our results highlight the importance of GW astronomy as a frontier in DM research, offering a complementary and potentially transformative approach to uncovering the nature of DM.

\section{Acknowledgments}
This work is supported by the National Natural Science Foundation of China (NNSFC) under Grant No.12475111, No.12205387, 
and the Fundamental Research Funds for the Central Universities, Sun Yat-sen University.


\begin{thebibliography}{99}
	
\bibitem{Jiang:2025blz}
S.~Jiang and F.~P.~Huang,
JCAP \textbf{06} (2025), 023
doi:10.1088/1475-7516/2025/06/023
[arXiv:2503.14332 [hep-ph]].


\bibitem{Huang:2025hos}
F.~P.~Huang, C.~Idegawa and A.~Yang,
[arXiv:2510.24007 [hep-ph]].
	
	
	
\bibitem{Huang:2017kzu}
F.~P.~Huang and C.~S.~Li,
Phys. Rev. D \textbf{96} (2017) no.9, 095028
doi:10.1103/PhysRevD.96.095028
[arXiv:1709.09691 [hep-ph]].
	
	
\bibitem{Jiang:2023nkj}
S.~Jiang, F.~P.~Huang and C.~S.~Li,
Phys. Rev. D \textbf{108} (2023) no.6, 063508
doi:10.1103/PhysRevD.108.063508
[arXiv:2305.02218 [hep-ph]].





\bibitem{Jiang:2023qbm}
S.~Jiang, A.~Yang, J.~Ma and F.~P.~Huang,
Class. Quant. Grav. \textbf{41} (2024) no.6, 065009
doi:10.1088/1361-6382/ad24c6
[arXiv:2306.17827 [hep-ph]].


\bibitem{Jiang:2024zrb}
S.~Jiang, F.~P.~Huang and P.~Ko,
JHEP \textbf{07} (2024), 053
doi:10.1007/JHEP07(2024)053
[arXiv:2404.16509 [hep-ph]].





\bibitem{Qiu:2025tmn}
D.~Qiu, S.~Jiang and F.~P.~Huang,
[arXiv:2508.04314 [hep-ph]].





\bibitem{Jiang:2025xln}
S.~Jiang, A.~Yang and F.~P.~Huang,
[arXiv:2511.23263 [astro-ph.HE]].


\end{thebibliography}
\end{document}